\documentclass[fleqn,twoside]{article}
\usepackage{espcrc2}
\usepackage{graphicx}
\title{R \& D for Future ZEPLIN}
\author{R. Bisset\address[Texas]{Department of Physics, Texas A\&M University,
    College Station, TX 77843, USA},
  M.J.Carson\address[Sheffield]{Department of Physics \& Astronomy, University
    of Sheffield, Hounsfield Road, Sheffield S3 7RH, UK},
  H. Chagani\addressmark[Sheffield],
  D.B. Cline\address[UCLA]{Department of Physics \& Astronomy, University of
    California, Los Angeles, CA 90095, USA},
  E.J. Daw\addressmark[Sheffield],
  T. Ferbel\address[Rochester]{Department of Physics \& Astronomy, Rochester
    University, Rochester, NY 14627, USA},
  J. Gao\addressmark[Texas],
  Y.S. Gao\address{Southern Methodist University, Dallas, TX 75275, USA},
  V.A. Kudryavtsev\addressmark[Sheffield],
  P.K. Lightfoot\addressmark[Sheffield],
  P. Majewski\addressmark[Sheffield]\thanks{Corresponding author, email p.majewski@sheffield.ac.uk},
  J. Maxin\addressmark[Texas],
  J. Miller\addressmark[Texas],
  W.C. Ooi\addressmark[UCLA],
  M. Robinson\addressmark[Sheffield],
  G. Salinas\addressmark[Texas],
  U. Schroeder\address{Department of Chemistry, Rochester University,
    Rochester, NY 14627, USA},
  J. Seifert\addressmark[Texas],
  F. Sergiampietri\addressmark[UCLA] \address{INFN Pisa, via Livornese 1291,
    San Piero a Grado (PI), Italy},
  W. Skulski\addressmark[Rochester],
  P.F. Smith\addressmark[UCLA] \address{Particle Physics Department,
    Rutherford, Appleton Laboratory, Chilton, Oxfordshire OX11 0QX, UK},
  N.J.C. Spooner\addressmark[Sheffield],
  J. Toke\addressmark[Rochester],
  H. Wang\addressmark[UCLA],
  J.T. White\addressmark[Texas],
  F. Wolfs\addressmark[Rochester],
  X. Yang\addressmark[UCLA]}
\begin{document}
\begin{abstract}
We propose a new concept for a very low background multi-ton liquid xenon
Dark Matter experiment. The detector consists of two concentric spheres and a
charge readout device in the centre. Xenon between the two spheres forms a
self-shield and veto device. The inner surface of the central sphere is coated
with CsI to form an internal photocathode with minimum of $2\pi$ coverage for
any event in the active volume. Photoelectrons from the CsI photocathode drift
toward the charge readout micro-structure in the centre of the detector. Both
scintillation and ionisation are measured simultaneously for background
rejection and 3-D event mapping. In addition to external shielding, the low
background is achieved by eliminating PMTs and by using low radioactivity pure
materials throughout the detector. We present detailed calculations of the
charge readout system and design details. The detector is expected to probe the
full SUSY parameter space.
\end{abstract}
\maketitle
\section{Introduction}
Liquid xenon is a promising target material for direct WIMP detection
experiments, due to its relatively high density of 3~$\mbox{g/cm}^{3}$,
allowing greater sensitivities to be reached with lower volumes. Additionally,
it has excellent ionisation and scintillation properties~\cite{Doke}.
Experimental programs that use liquid xenon as a target medium are currently
being pursued by the ZEPLIN~\cite{ZEPLIN}, DAMA~\cite{DAMALXe},
XMASS~\cite{XMASS} and XENON~\cite{XENON} Collaborations. To reach even greater
sensitivities than those proposed by the above experiments, a large target mass
and materials with a low intrinsic radioactive background need to be used. For
efficient scintillation light collection a multi-tonne detector would require a
significant number of photomultipliers to cover the large target surface. 
Compared to other sources of radioactivity, such as $^{85}\mbox{Kr}$
concentrations in xenon and ultra-pure copper, low background PMTs still
contribute significantly to the 
background, and hence affect the sensitivity of the detector~\cite{Carsonetal}.
An attractive alternative to PMTs is the CsI photocathode coupled with a
segmented charge amplifying and readout device. The low cost, flexible readout
configuration and relatively low intrinsic background of these devices with
respect to PMTs makes research in this area essential for the next generation
of dark matter detectors. Additionally, the use of liquid xenon as an active
veto surrounding the main target, means that fewer materials are required to
separate the two regions and 4$\pi$ coverage can be attained. In this paper the
feasibility of such a detector is discussed.
\section{Spherical Detector Principles and Operation}
A 10~cm radius ball that is covered with a charge collecting and amplifying
readout microstructure is located at the centre of the detector.
The LXe target with a mass of 1.3~T, is encased within a 1~cm thick
electroformed,  high-purity copper shell with a radius of 50~cm. The inner
surface of this shell is coated with CsI, acting as the photocathode. The
self-shield and veto consists of a sphere of liquid xenon surrounding this
copper shell, extending
to a radius of 81~cm from the centre of the detector. The entire structure is
insulated within a copper vacuum jacket. Field shaping rings are mounted on
insulator PTFE, which is attached to a
copper cylinder extending from the vacuum jacket to the central ball. The
signal is readout from anodes at ground. An additional electrode of the
microstructure 
is kept at negative potential, decoupling the drift and amplification regions.
This creates in a very short distance a very high electric field for charge
amplification in liquid.

An interaction in the target causes a simultaneous creation of scintillation
light and ionisation charge. The negative potential of the central sphere with
respect to the inner ball results in the radial drifting of electrons toward
the centre of the detector, where the amplification and signal 
readout take place. With highly segmented readout, ionisation electrons due to
short range 
of radiation tracks would produce a signal in a small number of
readout channels (primary pulse). In contrast, photons from scintillation are
emitted isotropically, converted into photoelectrons from CsI with a quantum
efficiency of
30\%~\cite{Aprileetal2}. These would produce a signal in a
larger number of readout channels (secondary pulse). Hence the signal from
charge and light can be additionally to its sequence better distinguished.
\begin{figure}[t]
  \begin{center}
    \includegraphics[scale=0.32]{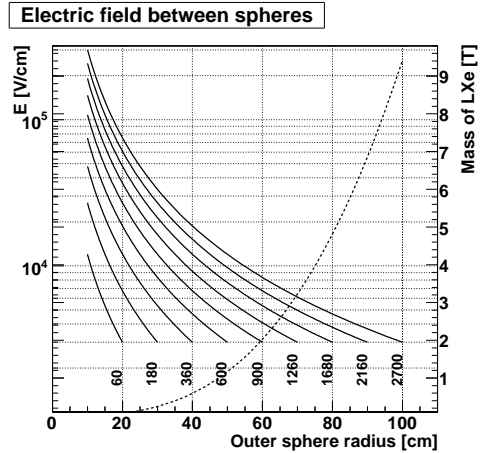}
    \vspace{-6mm}
    \caption{Electric Field between Spheres assuming that radius of inner ball
      is 10~cm. For an outer sphere radius of 50~cm, a potential difference of
      600~kV between the outer sphere and inner ball is required in order to
      maintain an electric field strength greater than 3~kV/cm across the
      target. The field at the surface of the inner ball is 75~kV/cm. The
      active target mass is approximately 1.3~T.}
    \label{field}
  \end{center}
  \vspace{-8mm}
\end{figure}
The time between these two pulses multiplied by the electron drift velocity
gives the radial coordinate information of the event. Other two coordinates are
delivered
from the position of the fired readout channels. It is essential that the
electrons drift at the same velocity, irrespective of the
non-uniform field across the liquid. However, as shown in figure~\ref{field},
the spherical detector operates with a non-uniform
field where $E\propto\frac{1}{r}$. The drift velocity of electrons in liquid
xenon saturates at 3~kV/cm~\cite{Milleretal}, hence for a central ball and
photocathode of radius 10 and 50 cm, respectively, a potential difference of
600~kV is required to create a minimum field in the chamber of 3~kV/cm~at the
photocathode, as shown in figure~\ref{field}. 
Additionally the quantum efficiency of the CsI gets higher in strong electric 
fields~\cite{Aprileetal2}.
\section{Charge Readout and Signal Feedback Problem}
\begin{figure}
  \begin{center}
    \includegraphics[scale=0.32]{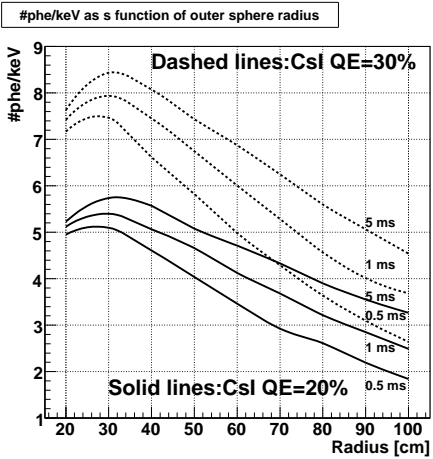}
    \vspace{-2mm}
    \caption{Results from Monte Carlo Calculations of the Number of
      Photoelectrons/keV as a Function of the Outer Sphere Radius. Electron
      lifetime values of 0.5, 1~and 5~ms are shown, for CsI QE=20 and 30\%. 
      With an outer sphere of radius 50~cm, a light collection efficiency
      varies between 4~and 7.5~photoelectrons/keV.}
    \label{MClight}
  \end{center}
  \vspace{-8mm}
\end{figure}

Good light collection is essential for detection of the secondary pulse.
Because light is converted into photoelectrons it depends on the purity of the
xenon, the attenuation length of light (100~cm in liquid
xenon~\cite{Baldinietal}) and the 
quantum efficiency of the CsI. The purity of liquid xenon is often quoted with
reference to the lifetime of electrons passing through it. Results from Monte
Carlo calculations, shown in figure~\ref{MClight}, indicate that a light
collection efficiency of 4~to 7.5~photoelectrons/keV is expected for an outer
sphere radius of 50~cm.
A threshold electric field strength of 1~MV/cm~\cite{Doke} is required for
avalanche developement, and hence charge amplification in liquid xenon.
Maximum gains of 100~\cite{Miyajimaetal} and 400~\cite{Derenzoetal} have been
observed in liquid xenon.
However, proportional scintillation light is created in electric fields greater
than 400~to 700~kV/cm~\cite{Masudaetal}, causing after-pulses and leading to
discharge. Other problems include local imperfections of the readout surface,
causing very high electric fields, and the slow motion of avalanche ions, thus
building space charge. Additionally, the level of maximum gain increases with
the purity of xenon. There are two types of possible charge readout devices
that have the potential
to operate at such high electric field strengths with small differential
voltage: cold field emission devices; and micropattern detectors.
Cold field emission devices have already been used in liquid argon, but no gain
was seen due to bubble formation on the sharp edges of the readout electrode,
hence creating a conduction path and discharges~\cite{Kimetal}. These devices
have not been tested in liquid xenon, and if the formation of bubbles can be
halted, then their use becomes an attractive alternative.
Conventional charge readout devices, such as Micromegas and Microstrip Gas
Chamber (MSGC) can also be used. A gain of 10 has already been observed with
MSGCs in liquid xenon~\cite{Policarpoetal}.

As discussed earlier, a non-ending cycle of after-pulses are seen due to the
threshold electric field strength for avalanche development being greater than
that for porportional scintillation light creation in liquid xenon.
A high electric field local to the charge readout device is essential for high
gain, as is a 100\% 4$\pi$ charge collection efficiency. One possible solution
is to use a high-voltage switch~\cite{Aprileetal4}, such that when the
potential of the cathode is 0~V, the maximum electric field strength local to
the charge readout device drops below 100~kV/cm preventing from proportional
light creation in LXe.  Another possible solution is
to use a light blocking focusing-defocusing device allowing electrons to travel
through to the
inner ball~\cite{ZEPLINIV}.
\section{Shielding}
An attractive feature of the detector geometry is the self-shielding provided
by a 30~cm thick layer of liquid xenon that constitutes the outer sphere. There
is the possibility of placing a charge readout device outside this layer, hence
making this an active veto.
\begin{figure}[t]
  \begin{center}
    \includegraphics[scale=0.35]{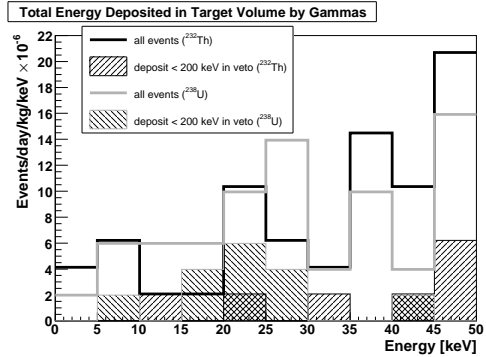}
     \vspace{-2mm}
    \caption{Total Energy Deposited in Target Volume by Gammas from
      $^{238}\mbox{U}$ and $^{232}\mbox{Th}$. The intensity drops from
      $0.504\pm 0.081$~to $0.525\pm 0.084$~events/day and $0.116\pm 0.039$~to
      $0.094\pm 0.034$~events/day for gamma-rays from $^{238}\mbox{U}$ and
      $^{232}\mbox{Th}$ respectively if an active liquid xenon veto is
      employed.}
    \label{shieldingplot}
  \end{center}
  \vspace{-8mm}
\end{figure}
Gamma-background simulations were performed with GEANT4~\cite{GEANT4}, for
$^{238}\mbox{U}$ and $^{232}\mbox{Th}$ content in the sphere holding this xenon
shield. A concentration of 0.05~ppm was assumed, which is the same as that in
current dark matter detectors. Over 30~cm, the number of interactions/day
dropped by over 2 orders of magnitude. Approximately $0.504\pm 0.081$~and
$0.525\pm 0.084$~events/day for interactions from $^{238}\mbox{U}$ and
$^{232}\mbox{Th}$ gamma-rays respectively was determined, as shown in
figure~\ref{shieldingplot}. Assuming that with an active veto, gamma-rays that
deposit greater than 200~keV in the shielding can be excluded, the respective
intensities drop to $0.116\pm 0.039$~and $0.094\pm 0.034$~events/day, also
shown in figure~\ref{shieldingplot}. Therefore, the veto provides a very good
shield against external gamma-ray sources.
\section{Plans for R \& D Program}
Future plans include: the study of the scintillation properties of liquid xenon
at high electric fields (such as the scintillation light and charge yield),
the determination of an accurate value of the electric field threshold for
proportional light creation and  maximum charge multiplication in LXe using
micro-structure devices.
\section{Summary}
The concept of a new sperical detector geometry using charge readout devices
coupled with CsI photocathode has been presented. Greater sensitivities can be
reached with this detector than with current dark matter search experiments due
to the large target mass, high efficiency of the scintillation light
collection, low radioactive background materials used in its composition and
the self-shielding properties of a 30~cm liquid xenon veto with 4$\pi$
coverage. The development of charge readout devices in liquid xenon is still 
in its infancy, and further research and development is required in this area
in order to realise this detector.

\end{document}